\def\publicversion{1}
\documentclass{article}
\usepackage{iclr2026_conference,times}
\usepackage[T1]{fontenc}
\usepackage{booktabs}
\usepackage{array}
\usepackage{tabularx}
\usepackage{tikz}
\usepackage{amsmath,amssymb}
\usepackage{xcolor}
\usepackage{url}
\usepackage[colorlinks=true,linkcolor=blue,citecolor=blue]{hyperref}
\hypersetup{pdftitle={Memory--Skill Isomorphism: One Skill Carrier, Two Native Uses}}

\newcommand{\measured}[1]{#1}   
\usepackage{graphicx}
\makeatletter
\begingroup
  \catcode`\/=\active \catcode`\.=\active \catcode`\:=\active \catcode`\-=\active
  \gdef\ttbrk@set{%
    \def/{\char47\ifhmode\penalty\z@\fi}%
    \def.{\char46\ifhmode\penalty\z@\fi}%
    \def:{\char58\ifhmode\penalty\z@\fi}%
    \def-{\char45\ifhmode\penalty\z@\fi}%
  }%
  \gdef\ttbrk@cats{\catcode`\/=\active \catcode`\.=\active \catcode`\:=\active \catcode`\-=\active}%
\endgroup
\protected\def\texttt{\ifmmode\expandafter\ttbrk@mm\else\expandafter\ttbrk@tx\fi}
\def\ttbrk@tx{\begingroup\ttfamily\ttbrk@cats\ttbrk@set\@ttbrk}
\def\ttbrk@mm#1{\mbox{\begingroup\ttfamily\ttbrk@cats\ttbrk@set#1\endgroup}}
\def\@ttbrk#1{#1\endgroup}
\makeatother

\newif\ifanonsubmission
\ifdefined\publicversion
  \anonsubmissionfalse
  \iclrfinalcopy
\else
  \anonsubmissiontrue
\fi

\title{Memory--Skill Isomorphism:\\
One Skill Carrier, Two Native Uses}
\ifanonsubmission
  \author{Anonymous authors}
\else
  \author{Kang Ruiyuan\\ X32 Studio\\ \texttt{kang@x32studio.com}}
\fi
\date{}

\ifanonsubmission
\hypersetup{pdfauthor={Anonymous authors}}
\else
\hypersetup{pdfauthor={Kang Ruiyuan}}
\fi

\begin{document}
\maketitle

\begin{abstract}
Memory and skills are two principal ways to improve agents without changing
weights: memory carries prior experience, skills carry reusable procedures.
Wrapping both in stores, routers, retrieval, reflection, and update paths makes
the machinery needed to reuse knowledge grow with what is accumulated.

Part of this duplication need not be rebuilt: a Skill is
already a natural carrier for distilled memory. In our
implementation, the
memory component is itself a Skill: resident \texttt{description} holds hot
cues, on-demand \texttt{SKILL.md} a colder index and curation
policy, and \texttt{reference/*.md} files detailed memories for
on-demand reading or search (levels $L_0 \to L_1 \to L_2$). The deployment's governed \measured{1{,}024}-character description budget forces
the resident index to stay compressed. Memory and capability can thus share one
progressive-disclosure carrier and read mechanism; reflection evolves either
Skill. Their writes still fork:
historical evidence is appended, whereas current indexes and procedures are
rewritten and revalidated.

In one deployed system, the most complete identification is governance:
\measured{4} write entries exist, but only \measured{1/4} reaches the
settlement ledger. At the priced operating point, one L1 lesson-1 point
measurement adds \measured{1{,}313} first-turn tokens against a
\measured{1{,}462}-token baseline---tied at $k=1$
(\measured{1{,}313} versus \measured{1{,}365}), \measured{4{,}949} at $k=5$; session totals differ by only
$\measured{1.18}\times$ with overlapping last-turn ranges, and price
decomposition is unavailable. File-copy migration needs no harness integration
or restarts, but behavioral preservation is untested (double
floor). On one selected task using one model, exposing the lesson is
associated with fewer failures (\measured{0/8} or \measured{1/8} with the
lesson versus a shared non-concurrent \measured{6/6} historical floor,
unadjusted for multiplicity). The task was selected on
prior floor evidence, so this is a selected-task post-selection existence
signal, not a confirmatory rate. Resident and BM25 carriers
show no detected behavioral difference in two small comparisons.
These measurements motivate a candidate RSI design rule: one Skill carrier
family for memory and capability, a shared read side, and governed distinct
writes. Body delivery $h := P(D \mid A)$ is
uninstrumented in RQ1--RQ2, so the evidence evaluates a resident-index implementation; the
resident description alone may carry the effect.
\end{abstract}

\section{Introduction: one shared read carrier, two write semantics}
\label{msec:intro}

Memory and skills are two central components of capable agents: memory carries
durable facts and experience; skills carry reusable procedures and executable
workflows~\cite{memorsurvey,agentskills,voyager}.
Systems therefore build memory stores and skill libraries, each with its own
retrieval and routing. Self-improving agents place another
loop around both: Hermes, for example, combines curated persistent memory,
autonomous skill creation and improvement, and cross-session
retrieval~\cite{hermesagent}. Each component is useful.

\paragraph{Why the problem compounds.}
Separate subsystems imply separate stores, routing decisions, context injection, update APIs, and validators, so a successful loop makes its own control plane harder to operate. Collapsing everything into
one mutable memory is not the answer: rewriting current advice must not erase
the historical evidence that justified it. The first-principles question is
therefore: \emph{which operations genuinely differ across knowledge classes, and
which are duplicate machinery?}

\paragraph{The overlooked reuse.}
To change a future action, off-weight knowledge must remain in
context or become discoverable on demand. A standard Skill already implements
both~\cite{agentskills,anthropic-skills}: its short \texttt{description} $L_0$
is mounted proactively; a matching task opens the body $L_1$; supporting
references $L_2$ are read or searched only when deeper detail is
needed---the hot, warm, and cold read requirements of distilled experience.
We therefore make $\phi$ literal: the memory component is not a service beside
the Skill system; it \emph{is a memory Skill}. Its $L_0$
stores the hottest cues, its $L_1$ stores a colder memory index and curation
policy, and its $L_2$ stores detailed lessons; Table~\ref{tab:memory-skill-map} gives the complete mapping.

\begin{figure}[t]\centering
\resizebox{0.93\textwidth}{!}{\input{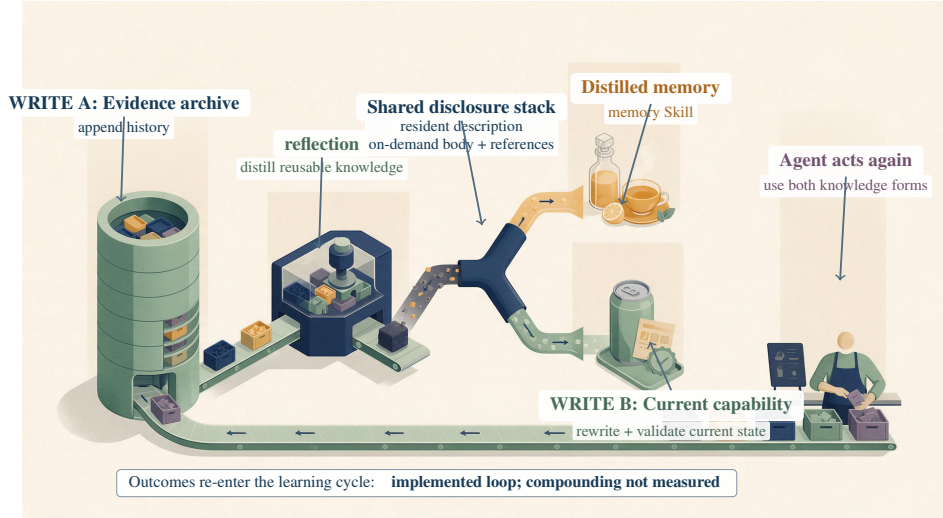}}
\caption{From outcomes to reusable knowledge. The memory Skill represents
distilled experience as a Skill: resident description $\rightarrow$ on-demand
body $\rightarrow$ searchable references. Capability Skills reuse the same
progressive-disclosure carrier, so reflection evolves one carrier family.
WRITE A accumulates history; WRITE B rewrites and validates current state.
Arrows show the implemented loop structure (backdrop AI-generated; labels
overlaid); compounding gain remains unmeasured; body delivery is unmeasured in RQ1--RQ2 and instrumented in E1 (\S\ref{sec:e1}); write audit:
\measured{4} entries, ledger \measured{1/4}.}
\label{fig:carrier}
\end{figure}

This observation organizes reflection around one decision: classify new evidence by semantics---archive raw outcomes, update the memory Skill with reusable experience and capability Skills with executable procedure---reusing one carrier family, mounting, loading, and structural validation. The reuse stops at writing: history is append-only while current indexes and procedures must be rewritable and revalidated. Our thesis is therefore:
\textbf{memory and skill are isomorphic (definitionally, by construction) at the
progressive-disclosure carrier
($\phi$ preserves reads; \S\ref{msec:formal}), while safe RSI must preserve two
write semantics}
(append $H_{t+1} = \mathrm{append}(H_t, e)$ versus rewrite--revalidate
$S_{t+1} = \mathrm{validate}(f(S_t))$). The body-delivery half would fall if E1's description-only
arm succeeds alongside full disclosure (preregistered sufficiency zone).
Figure~\ref{fig:carrier} shows the resulting loop.

We evaluate the thesis as a single-deployment audit with a priced
operating-point note, separating availability ($P(\mathrm{fail} \mid \mathrm{lesson}) \in
\{\measured{0/8}, \measured{1/8}\}$ versus $P(\mathrm{fail} \mid \neg\mathrm{lesson})
= \measured{6/6}$ on the selected task) from carrier form (two nulls:
$\measured{8/10}$ vs $\measured{8/10}$ and $\measured{6/8}$ vs $\measured{5/8}$,
not equivalence). First-turn $C_{\mathrm{res}} = \measured{1{,}313}$ tokens versus
$C_{\mathrm{ret}}(1) = \measured{1{,}365}$ and $C_{\mathrm{ret}}(5) = \measured{4{,}949}$---with
session prompts at $\measured{1.18}\times$ and overlapping last-turn ranges;
$\measured{0}$ of $\measured{11}$ content changes touch harness code against
$\measured{15}$ machinery edits; ledger coverage is
$\measured{1/4}$. These establish resource and governance properties,
not behavioral superiority.

\paragraph{Contributions.}
\textbf{(1) Memory--Skill isomorphism (a definitional carrier mapping, not a
measured advantage):} the map $\phi$ sends proactive
mounting, on-demand reading, progressive disclosure, bounded resident cues,
and deep lookup onto native Skill structure ($L_0 \to L_1 \to L_2$), making
memory one concrete Skill rather than a parallel memory service.
\textbf{(2) A shared-carrier evolution design:} reflection evolves experience
and capability through one carrier family, while append-only history
($H_{t+1} = \mathrm{append}(H_t, e)$) and rewritable current state ($S_{t+1} =
\mathrm{validate}(f(S_t))$) retain different write semantics. \textbf{(3) An
operating-point measurement:} behavior, context, migration, and write coverage
jointly locate where this reuse lowers first-turn context and
where governance remains incomplete: \measured{4} write entries
with ledger coverage $\measured{1/4}$.

\section{Related work}
\label{msec:related}

\paragraph{Memory and skill carriers.}
Agent memory spans retrievable logs, streams, banks, and external stores
\cite{generativeagents,memorybank,memgpt,mem0,amem}; benchmarks continue to expose disagreements about
what long-term memory should measure~\cite{longmemeval,membench}. Skills are commonly treated as
procedural memory~\cite{memorsurvey}. Agent Skills supplies the trigger--body
format and progressive disclosure used here, while Voyager already pairs
descriptions with on-demand skill bodies~\cite{agentskills,anthropic-skills,voyager}.
Agent tooling also routes durable instructions through memory and skill
files~\cite{claudecode-memory}. MemGPT's core versus archival memory already
separates mutable resident state from append-oriented external
history~\cite{memgpt}; event sourcing established append-only logs with
materialized current views~\cite{fowleventsourcing}. These
are format, practice, and state-management precedents. We study a different
boundary: where this fork lands on a shared Skill disclosure stack, and
whether a deployment actually governs every mutation path.

\paragraph{Evolving textual artifacts.}
ExpeL distills experience, and Agent Workflow Memory induces reusable
procedures from trajectories~\cite{expel,agentworkflowmemory}. Dynamic
Cheatsheet, ACE, and Symbolic Learning evolve textual state, while Reflexion
and Self-Refine use language feedback within or across
episodes~\cite{cheatsheet,ace,symboliclearning,reflexion,selfrefine}. Hermes runs
a learning loop over distinct
memory and skill facilities, while Memento-Skills explicitly evolves
structured skill files as persistent memory~\cite{hermesagent,mementoskills}.
We therefore do not claim the first reflection loop, self-authored skill, or
skill-as-memory usage. Our delta is to map memory requirements to the native
Skill disclosure stack, measure that reuse against a separate retriever, and
audit the write semantics and governance it leaves behind.

\paragraph{Self-improvement and retrieval.}
Self-improvement spans self-evolving systems and weight-level adaptation~\cite{godelmachine,selfevolving,seal};
retrieval augmentation provides the external-store alternative~\cite{rag,dpr}.
Our comparison instead measures first-turn cost and governance for a compact set of distilled lessons. Recursive self-improvement (RSI) studies such loops directly~\cite{rsi-survey}; this paper contributes one audited iteration on a shared carrier.

\paragraph{Gap.}
Prior systems establish both separate memory/skill facilities and evolving
skills as memory, and we concede the format, the loop, and skill-as-memory
usage. What remains missing is a measurement-and-governance account for a
shared Skill carrier: which disclosure operations distilled memory reuses,
what that reuse costs against a separate retriever, and where write governance
diverges. We supply that account under one deployment boundary,
organized by Eq.~\eqref{eq:cascade} and Design Rule~1.

\section{Design: one read carrier, two write semantics}
\label{msec:method}

Knowledge that never re-enters the decision context cannot change an action.
Off-weight knowledge returns by staying resident or being pulled, and pulling
needs a resident trigger or an external retriever~\cite{seal,rag,selfrag,mem0}.
The design question is: \emph{what is worth keeping resident?}

\subsection{Formal setup}
\label{msec:formal}

A Skill $s$ exposes three disclosure levels: $L_0(s)$, the resident
description; $L_1(s)$, the body opened on demand; and $L_2(s)$, references
read or searched on demand. Two tiers share this shape: distilled memory $m$
and capability $c$. Read operations are identical for both:
\begin{equation}
\mathrm{mount}(L_0), \quad \mathrm{open}(L_1 \mid L_0), \quad
\mathrm{query}(L_2 \mid L_0).
\label{eq:reads}
\end{equation}
Let $\phi: \mathcal{N} \to \mathcal{S}$ map the five memory read needs
$\mathcal{N}$ onto the five Skill affordances $\mathcal{S}$ (the rows of
Table~\ref{tab:memory-skill-map}), bijectively by construction. It preserves
reads and pointedly fails on writes: the history log
$H_t$ appends,
\begin{equation}
H_{t+1} = \mathrm{append}(H_t, e),
\label{eq:append}
\end{equation}
while current state $S_t$ rewrites and revalidates, for system-specific
procedures $f$ and $\mathrm{validate}$ (on rejection, $S_{t+1} = S_t$),
\begin{equation}
S_{t+1} = \mathrm{validate}(f(S_t)).
\label{eq:rewrite}
\end{equation}
The ``isomorphism'' of the subtitle is exactly this bijection $\phi$ on reads, in
the finite sense: five read requirements matched to five Skill affordances. It
asserts no shared write semantics, content, provenance, or lifecycle, and
$\phi$ admits no extension to writes that would identify append with rewrite.

A lesson's path to behavior travels $A \to D \to U \to O$: availability $A$,
body delivery $D$, use $U$, and the observable task endpoint $O$. The chain
below assumes Markov dependence (each stage only on its predecessor). The
chain path is not necessary---all four lesson-3 units apply without $D$---so
it governs lesson-caused effects only:
\begin{equation}
P(A, D, U, O) = P(A)\,P(D \mid A)\,P(U \mid D)\,P(O \mid U).
\label{eq:cascade}
\end{equation}
Causal readings come only from controlled contrasts---here, RQ1's
post-selected floor arm---never from $O$ alone. $U$ and $O$ currently
share the frozen-checker proxy; separating them needs a process trace. The
delivery probability $h := P(D \mid A)$ is the common missing instrument:
without it, a null outcome cannot separate routing failure from a weak lesson.

First-turn knowledge cost is the measured prompt delta against the L1 $k{=}0$
baseline ($1{,}462$ tokens), before any body opens:
\begin{equation}
C_{\mathrm{res}} := \Delta^{(1)}_I, \qquad C_{\mathrm{ret}}(k) := \Delta^{(1)}_R(k),
\label{eq:cost}
\end{equation}
evaluated once (Table~\ref{tab:runtime-cost}): $C_{\mathrm{res}} =
\measured{1{,}313}$ from behavioral-batch lesson-1 $I$ units;
$C_{\mathrm{ret}}(1) = \measured{1{,}365}$, $C_{\mathrm{ret}}(3) =
\measured{3{,}980}$, $C_{\mathrm{ret}}(5) = \measured{4{,}949}$ from the
lesson-1 R-arm token curve. Structurally this realizes mounted descriptions
plus six-skill scaffolding versus injected blocks, but the equation names the
measured deltas, not the decomposition. Sessions need their own priced sum;
dollars are not Eq.~\eqref{eq:cost} times turns.

These definitions support one compact design rule:

\begin{quote}\small
\textbf{Design Rule 1 (shared carrier).} Two distilled tiers may share one
disclosure stack when (i) both re-enter decisions through a trigger--body
shape, (ii) description routing is an assumption to be instrumented, not a
measured mechanism, and (iii) each tier declares
write invariants the append/rewrite fork preserves. The rule is refuted by
incompatible discovery operations or by one governed transition covering both
invariants (see Falsifiability). As a consistency check (not an independent
test), the rule agrees with the design on the raw archive, which fails
(ii)--(iii) and stays unmounted. E1's arms $D_0$ (description-only) and $D_1$
(full resident) probe (ii) as a future test: only delivery-stratified gains
with $h$ instrumented could support the shared carrier; the lesson-2/3
dissociations already refute a trigger--body causal reading here.
\end{quote}

\subsection{Why memory can be a Skill}
\label{msec:physics}

The memory component is not a service beside the Skill system. It is a Skill
directory loaded by the same mechanism as every capability Skill. The
shared read structure follows from three levels of disclosure. The host proactively
mounts each \texttt{description} $L_0$ as a resident trigger. When that trigger
is relevant, the agent can open the body $L_1$; it reads or searches
references $L_2$ only when the task needs deeper detail. In the memory
Skill, these levels become a hot memory index, a colder index plus memory
curation policy, and detailed cold memories. The map $\phi$ reuses
mounting, routing, loading, and supporting-file lookup rather than building a
second memory retriever. The format itself remains prior
art~\cite{agentskills,anthropic-skills}.

The component constraints make the reuse consequential. A controlled-write
gate caps each resident \texttt{description} at
\measured{1{,}024} characters. Because that description text recurs in every
request, the cap forces the hot experience index to be compressed and rewritten as
current state while the evidence and reference store remains uncapped. It
bounds the resident-text \emph{gateway} without guaranteeing that the
model attends to a trigger or pulls the correct body. The latter is the
body-pull fraction $h$, unmeasured in RQ1--RQ2 and instrumented in E1 (\S\ref{sec:e1}).

\begin{table}[t]
\centering\small
\caption{Distilled memory requirements map onto native Skill
affordances (the 1,024-character limit belongs to the governed deployment, not
the Agent Skills specification).}
\label{tab:memory-skill-map}
\begin{tabularx}{\textwidth}{@{}p{2.35cm}p{3.05cm}X@{}}
\toprule
Memory requirement & Native Skill affordance & Memory realization \\
\midrule
Proactive mounting & Resident \texttt{description} ($L_0$) & Hot cues and the
highest-value memory index \\
On-demand reading & Body $L_1$ opened when relevant & Colder index plus
memory and curation policy \\
Progressive disclosure & $L_0 \rightarrow L_1 \rightarrow L_2$ &
Hot cues $\rightarrow$ cold index $\rightarrow$ detailed memory \\
Bounded resident gateway & Governed \measured{1{,}024}-character description &
Reflection must compress or replace cues rather than accumulate them \\
Deep lookup & Supporting files can be read or searched & Detailed
\texttt{reference/*.md} memories, including grep-style lookup \\
\bottomrule
\end{tabularx}
\end{table}

Not every memory object belongs in this stack. Raw outcomes and provenance are
source material rather than distilled policy; they remain in an unmounted,
append-only archive and are searched during reflection.

\subsection{The improvement loop}

Figure~\ref{fig:carrier} shows how the parts compose. This loop instantiates
the reflector--curator pattern studied in prior experience-learning
systems~\cite{ace,expel}; the contribution here is its carrier and governance,
not the loop itself. Task outcomes first remain as attributable evidence before
reflection routes them: raw observations stay archived, situational lessons
update the memory Skill, and stable procedures revise a capability Skill, with
only the two distilled tiers sharing one discovery loader. On later tasks,
$m$ and $c$ shape new actions, producing new
evidence, closing the recursive feedback path from outcomes $O$ back to
availability $A$; compounding gain across cycles is not yet measured, and no
saving against a separately built memory subsystem is claimed here. Update
invariants stay distinct, with history accumulating while current indexes and
procedures are replaced under validation.

\ifanonsubmission
[companion repository withheld for review].
\else
A companion repository publishes the same organizational guidance and
agent-facing skill set~\cite{x32skillsystem}.
\fi

\subsection{Why the write layer must fork}
\label{msec:surface}

Sharing a read carrier does not imply one write operation. Historical evidence
must preserve provenance, so it is appended. Current procedures must track the
best known state, so they are replaced and then revalidated. Both are intended
to obey one governance contract while retaining different
state transitions; measured ledger coverage is \measured{1/4} (Table
\ref{tab:write-audit-main}), so unification is a goal, not a result. Here, the
settlement ledger pairs a predicted consequence with the observed post-write outcome.

\begin{table}[t]
\centering\small
\caption{Read sharing and write governance in the audited deployment.}
\label{tab:write-audit-main}
\begin{tabularx}{\textwidth}{@{}p{3.15cm}p{2.0cm}X@{}}
\toprule
Audit question & Observed & Interpretation \\
\midrule
Do distilled tiers share a read carrier? & Yes & Same trigger--body discovery
shape \\
How many write entries were enumerated? & \measured{4} & A single physical
path is refuted in this deployment \\
How many enumerated entries reach the ledger? & \measured{1/4} & Three observed
entries escape the common ledger \\
Must history and current state write differently? & Yes & Eqs.~\eqref{eq:append}--\eqref{eq:rewrite}
preserve different invariants \\
\bottomrule
\end{tabularx}
\end{table}

The audit changes the engineering claim: not ``one update
function'' but one auditable contract covering every official entry while
preserving append-only history and rewritable current state. The present
implementation has not reached that goal; an ordinary file-write bypass
remains outside the settlement ledger. Appendix
\ref{sec:audit} enumerates the paths, source locations, and contradictory
contract clauses. This is Design Rule~1 in prose: the audit tests whether the
deployed write paths honor the fork the rule requires.

\paragraph{Falsifiability.}
The write fork would be unnecessary if one governed transition could preserve
provenance while retiring obsolete procedure without distinguishable replacement
semantics; a single append-only log with validated current-state projections
still exposes both transitions, so it is an instance of this framework, not a
counterexample. No cross-implementation impossibility is claimed; the fork is
a deployment-level requirement.

\section{Evaluation}
\label{msec:experiments}

The shared-carrier audit earns its keep only if three measurements hold. The
knowledge must matter to behavior; the native Skill carrier must be priced
against a retrieval baseline rather than asserted by definition; and the design
goal is to avoid a second discovery loader (the subsystem comparison itself
was not run). Each question is answered within a single batch and model,
except the runtime table, split into point-measurement and batch panels with
per-row denominators; no cross-batch pooling is performed. Three questions
organize the measurements:

\begin{enumerate}
\item \textbf{Value:} does $P(\mathrm{fail} \mid \mathrm{lesson}) < P(\mathrm{fail} \mid \neg\mathrm{lesson})$?
\item \textbf{Carrier comparison:} how do lesson application, i.e.\ use $U$
per arm ($U^I$ vs $U^R$), and runtime cost ($C_{\mathrm{res}}$ vs
$C_{\mathrm{ret}}(k)$) vary with delivery form?
\item \textbf{RSI operation:} what integration work, maintenance coupling,
and ledger coverage remain after reuse?
\end{enumerate}

All behavioral decisions use frozen task checkers and paired seeds where
available; invalid and superseded runs remain archived outside the reported
denominators. Appendix~\ref{sec:pilot} contains the
protocol fingerprints, task chronology, and per-unit records.

\subsection{RQ1: An exploratory anecdote on knowledge availability (selected task)}
\label{msec:pilot}

On the selected delegation task with \texttt{opencode-go/deepseek-flash},
failure means omitting an explicit provider/model binding or the post-dispatch
runner confirmation (frozen conjunctive checker), and lower
reproduction is better. Lesson, model,
paired-seed design, and per-arm size were fixed before the expansion
run; the result remains specific to that selected lesson. With $F$ the failure
event and $\alpha, \beta, \gamma$ the arms, $P(F \mid \alpha) = \measured{0/8}$,
$P(F \mid \beta) = \measured{1/8}$, and $P(F \mid \gamma) = \measured{6/6}$ for
the historical floor (exact Fisher values in Appendix~\ref{sec:pilot}: a
historical, non-concurrent, shared-arm, unadjusted, post-selection
computation). Thus, for this task and model, exposing the lesson
is associated with different outcomes in this exploratory anecdote. These values
are post-selection descriptive statistics; they do not generalize to a
population of tasks. A preregistered description-only ablation (E1, \measured{n=8}/arm, paired seeds; \S\ref{sec:e1}) gives $P(F\mid D_0)=\measured{0/8}$ versus $P(F\mid D_1)=\measured{1/8}$ ($\Delta=\measured{-12.5}$\,pp, Fisher $\measured{p=1.0}$, CI $[\measured{-47.09},\measured{+30.20}]$\,pp, descriptive): the body adds nothing on this task.

The two forms serve primarily as independent realizations of knowledge-present;
their comparison is secondary ($\Delta = \measured{-12.5}$\,pp, CI
$[\measured{-47}, \measured{+30}]$\,pp, $\measured{p=1.0}$; paired McNemar
$\measured{p=1.0}$; the single failure misses only the confirmation step):
availability on this task and model, not phrasing or carrier. One arm carries
the rule at the resident level while both expose the full detail store, so the
batch cannot separate cue-only from full-body effects---the prerequisite holds,
not the mechanism.

\begin{table}[t]
\centering\small
\caption{Behavioral results answer distinct questions (failure lower-is-better;
application higher-is-better). Tests are Fisher
exact; exact availability $p$-values are in Appendix~\ref{sec:pilot}. A screened
rerun (\measured{6/8} vs \measured{5/8}) was marked insufficient/void and is
discussed in text only.}
\label{tab:behavior}
\begin{tabularx}{\textwidth}{@{}p{2.55cm}p{2.35cm}p{2.0cm}X@{}}
\toprule
Question & Contrast & Result & Conclusion \\
\midrule
Knowledge available? & Form A vs no lesson & \measured{0/8} vs
\measured{6/6} & Selected-task existence signal \\
Knowledge available? & Form B vs no lesson & \measured{1/8} vs
\measured{6/6} & Same directional signal \\
Writing form better? & A vs B & \measured{0/8} vs \measured{1/8};
\measured{p=1.0} & No detected difference \\
Carrier form better? & Resident vs BM25 & \measured{8/10} vs
\measured{8/10}; \measured{p=1.0} & No detected difference \\
\bottomrule
\end{tabularx}
\end{table}

\subsection{RQ2: Underpowered carrier comparison with lower first-turn context}
\label{msec:cost}

Arms are $I$ (resident index) and $R$ (BM25 retrieval): a deployed 6-skill
versus 0-skill contrast with unconditional sparse-BM25 injection only. The main carrier
comparison holds tasks, model (\texttt{openrouter/z-ai/glm-5.3-flash}), and
paired seeds fixed. Arm $I$ keeps a compact index in every request and reads
bodies on demand. The BM25
arm unconditionally injects the top-$k$ retrieved body text every turn; a
conditional policy would shift the profile (Appendix~\ref{sec:compare}). Table
\ref{tab:runtime-cost} reports the knowledge component relative to a
\measured{1{,}462}-token no-knowledge first-turn baseline.

\begin{table}[t]
\centering\small
\caption{Priced operating point, split by denominator. (a) First-turn tokens:
one L1 lesson-1 point measurement per cell, slice labeled per row. (b) The
$k=5$ behavioral batch (5 tasks, ten paired units per arm). Session prompts
$\measured{1.18}\times$; cache split $\measured{1.84}\times$ /
$\measured{1.05}\times$; price decomposition unavailable.}
\label{tab:runtime-cost}
\begin{tabularx}{\textwidth}{@{}p{2.35cm}p{2.0cm}X@{}}
\toprule
(a) Carrier & First-turn added tokens & Slice (differs by row) \\
\midrule
Resident index & \measured{1{,}313} & Behavioral batch, lesson-1 $I$ \\
BM25, $k=1$ & \measured{1{,}365} & R-arm token curve (point only) \\
BM25, $k=3$ & \measured{3{,}980} & Same curve (point only) \\
BM25, $k=5$ & \measured{4{,}949} & Same curve (point only) \\
\bottomrule
\end{tabularx}
\begin{tabularx}{\textwidth}{@{}p{2.5cm}p{2.5cm}X@{}}
\toprule
(b) Carrier & Applied lesson & Batch cost \\
\midrule
Resident index & \measured{8/10} & \measured{\$0.09619} \\
BM25, $k=5$ & \measured{8/10} & \measured{\$0.12521} \\
\bottomrule
\end{tabularx}
\end{table}

At $k=1$, $C_{\mathrm{res}} \approx C_{\mathrm{ret}}$; at $k=5$,
$C_{\mathrm{ret}} / C_{\mathrm{res}} = \measured{3.8}\times$ first-turn. In
the behavioral batch, provider-reported cost is $\measured{1.302}\times$
higher for retrieval, while $U^I = U^R = \measured{8/10}$. First-turn accounting
therefore favors the resident index at $k=5$ in this point measurement, not
overall or behavioral superiority: the wide uncertainty below still permits
consequential behavioral differences. The shared carrier avoids in-loop
retrieval calls at this operating point, but session-level prompt volume differs
by only $\measured{1.18}\times$ with overlapping last-turn ranges, so
first-turn deltas must not be read as session savings; knowledge-component
chars differ by $\measured{3.41}\times$
(\measured{268{,}576} vs \measured{916{,}903}); the comparison is underpowered
and cannot distinguish carrier forms.

Paired by lesson and seed, the ten pairs split $\measured{8}$ agree-pass,
$\measured{2}$ agree-fail, and $\measured{0}$ discordant (McNemar
$\measured{p=1.0}$, uninformative); the Fisher value and CI below are unpaired
descriptives. The uncertainty is wide: $\Delta = \measured{0}$\,pp with
CI $[\measured{-45.3}, \measured{+45.3}]$\,pp. A screened rerun likewise gives
$\measured{6/8}$ versus $\measured{5/8}$, CI
$[\measured{-35.5}, \measured{+57.5}]$\,pp, $\measured{p=1.0}$.
The archived recomputer marks the screened batch insufficient/void, so these
are descriptive values excluded from all confirmatory reading; the literal
preregistration clause would have read ``supported'', so voiding on the T2
exclusion was a post-hoc conservative override. Neither study
is an equivalence test.

The diagnostic \texttt{target\_found} values are deliberately not compared:
an observed reference-file read and inclusion in injected text answer different
questions. The body-pull fraction remains the common missing instrument.

Within the resident arm, observed detail reads occur in only
\measured{3/10} units while the failure is avoided in \measured{8/10}; the
retained cascade dissociations---retrieved lesson-2 units failing despite
delivery and lesson-3 units applying without it---already refute a
trigger--body causal reading here, so no routing-sufficiency mechanism is claimed.

Retrospective proxy tabulation finds no detected association in either arm
(all $\measured{p=1.0}$): resident \measured{3/3} vs
\measured{5/7}; retrieved \measured{6/8} vs \measured{2/2}; screened resident
\measured{2/2} vs \measured{4/6}---within-arm descriptives only, never
contrasted across arms. Both cascade dissociations appear, retained per
protocol; a prospective delivery instrument remains required
(Appendix~\ref{sec:pilot}).

\subsection{RQ3: The carrier moves without harness integration; governance remains}

\begin{table}[t]
\centering\small
\caption{Migration, maintenance, and write-governance measurements.}
\label{tab:portability-main}
\begin{tabularx}{\textwidth}{@{}p{3.2cm}p{2.15cm}X@{}}
\toprule
Measurement & Result & Conclusion \\
\midrule
Migration start snapshot & \measured{21} files,
\measured{177{,}321} bytes & File-level portability; second snapshot disagrees \\
Integration work & \measured{0} code lines,
\measured{0} restarts & Copy, not harness integration \\
Behavior after migration & \measured{0/8} vs \measured{0/8},
\measured{p=1.0} & Double floor; preservation untested \\
Content-change window & \measured{+64\%} store growth,
\measured{0/11} harness touches & Content additions were decoupled in this
window \\
Machinery changes & \measured{15} edits & Building and maintaining machinery
was not free \\
\bottomrule
\end{tabularx}
\end{table}

The carrier moves cleanly as files, but preservation is untestable at double
floor. Likewise, eleven content changes touched no harness code, yet the
machinery itself changed fifteen times (\measured{13} instrumentation plus
\measured{2} write-entry): content addition decoupled from code in one window,
not zero maintenance. Two payload snapshots disagree (\measured{177{,}321} vs
\measured{185{,}582} bytes; Appendix), so the conclusion is copy-based
integration, not byte-identical carriers.

Together, these measurements locate the remaining problem: read-side sharing
lowers first-turn tokens in this point measurement with copy-level integration,
but write-side consolidation is incomplete, with
mutation paths still escaping the common settlement contract.

\paragraph{Boundaries.}
The study covers one deployment and small behavioral samples;
Appendix~\ref{sec:boundaries} gives the complete claim contract and
protocol-specific qualifications.

\section{Discussion: orthogonal axes and the operating point}
\label{msec:discussion}

\paragraph{Availability, carrier form, and write semantics are different axes.}
The experiments become coherent once these axes are separated. Availability asks
whether the task-relevant lesson is exposed ($A$ assigned); the floor
comparison gives a selected-task existence signal. Carrier form asks whether fixed
$C_{\mathrm{res}}$ undercuts dynamic $C_{\mathrm{ret}}(k)$ in this point
measurement, at indistinguishable \measured{8/10} vs \measured{8/10} whose
interval still permits $\sim$45\,pp. Write semantics asks which state transition is
valid: Eq.~\eqref{eq:append} or Eq.~\eqref{eq:rewrite}. That axis is established
by artifact invariants and audited entry points, not by the behavioral
comparison. A carrier-form null does not erase the availability result,
and availability does not prove either carrier superior.

\subsection{A measurement stack for evolving agents}

RAG evaluation already separates retrieval from support~\cite{rag}; with delivery instrumented, a null still locates routing, application, or task execution. Eq.~\eqref{eq:cascade}
models the lesson-caused path alone; outside that path is
off-model by design, and endpoint-only testing cannot distinguish a weak lesson from routing failure.

\begin{table}[t]
\centering\small
\caption{Four loop measurements. Availability and outcome are identified on one
selected task with an application proxy; delivery is instrumented only in E1.}
\label{tab:measurement-stack}
\begin{tabularx}{\textwidth}{@{}p{2.3cm}p{3.25cm}X@{}}
\toprule
Stage & Question & Status in this study \\
\midrule
Availability & Is the relevant lesson exposed to the agent? & Controlled by
the knowledge-present versus floor arms \\
Delivery & Did the intended body enter context before action? & Unmeasured
common quantity $h$; confounded with availability in RQ1 \\
Use & Did the artifact avoid the recorded failure? & Frozen checker; an
application proxy, not a process trace \\
Outcome & Did task behavior change? & Selected-task existence signal;
carrier comparison underpowered \\
\bottomrule
\end{tabularx}
\end{table}

\paragraph{The shared carrier has a specific operating point.}
Table~\ref{tab:runtime-cost}
differs across the measured points: near-tied at single-body retrieval,
resident smaller at broader retrieval. Growing carrier sets, semantic-match misses, and cache pricing can move the comparison; no tool-call interface appears in all \measured{75} rows carrying the field. Prospective delivery logging in E1 records zero body/reference opens in \measured{16/16} units, closing the uninstrumented-h gap with a zero reading rather than a new gap.

\paragraph{Governance determines whether consolidation stays safe.}
Sharing discovery moves responsibility to the shared interface, so ledger coverage is the deployment criterion: every official entry must
participate without erasing either transition.

\paragraph{What should be measured next.}
The missing instrument was delivery: E1 (\S\ref{sec:e1}) records whether each matched body entered
context before action on the L5 task and reads zero (\measured{0/16} opens); repeating it at larger $n$
across tasks is next. The observed null claims
no routing-sufficiency mechanism: the retained lesson-2/3 dissociations refute a trigger--body causal reading.

\paragraph{An RSI reading.} One loop iteration is audited end to end---experience as attributable evidence, reflection distilling it, the skill carrying it back, the entry-point audit bounding rewrites. E1 adds zero body opens (\measured{0/16}) on the L5 task. Genuine RSI further requires compounding across cycles (unmeasured) and ledger coverage of every write path (\measured{1/4} here): the carrier is a candidate substrate, not a demonstration.

\section{Conclusion}
\label{msec:conclusion}

Memory and skill need not imply two read systems: memory is a Skill whose
$L_0 \to L_1 \to L_2$ disclosure implements
hot-to-cold memory access. Executable capability uses the same carrier, so
reflection evolves one Skill family for knowledge and capability; \textbf{One Skill carrier, two native uses} is the read-side result
(Eq.~\eqref{eq:reads}). Safe evolution adds the fork:
Eq.~\eqref{eq:append} for history, Eq.~\eqref{eq:rewrite} for current state.

The experiments distinguish value from mechanism. Relevant distilled knowledge
is associated with fewer failures on the selected single-model task in an
exploratory anecdote, while the underpowered carrier comparison cannot
distinguish forms (no detected difference in either small comparison). The
structural evidence is independent: lower first-turn context in a point
measurement, copy-based migration, and harness-decoupled content additions.

Sharing a carrier does not automatically unify
governance: the audit finds \measured{4} write entries with only \measured{1/4}
reaching settlement. Next: instrument body delivery
before enlarging another carrier-form batch; closing the three
unledgered entries stays open. E1 is reported as Z-mixed (provisional) under its frozen rule: an O-difference with no carrier conclusion.
The present evidence supports
selected-task knowledge
availability, first-turn context accounting, windowed maintenance decoupling,
and separate write semantics---not behavioral superiority: a single-deployment
audit plus a priced operating-point note, with every number linked to
its source in the appendix.

\clearpage
\bibliographystyle{iclr2026_conference}
\ifanonsubmission
\bibliography{refs-v2}
\else
\bibliography{refs-v2,refs-v2-public}
\fi

\clearpage
\appendix

\section{Evidence contract and artifact map}
\label{sec:boundaries}

The paper reports a deployed case study, not a universal benchmark. Table
\ref{tab:appendix-contract} records the claim boundary that governs every
result. Task prompts, unit records, outputs, checker code, source excerpts,
and hashes are distributed with the artifact; session transcripts and the
external deployment history are not. The appendix summarizes rather than
reproduces those files.

\begin{table}[h]
\centering\small
\caption{Claim contract. A null superiority test is never interpreted as
equivalence.}
\label{tab:appendix-contract}
\begin{tabularx}{\textwidth}{@{}p{2.6cm}p{3.0cm}X@{}}
\toprule
Question & Evidence & Supported conclusion \\
\midrule
Knowledge availability & Selected L5 floor contrast & Post-selection
descriptive association on this task and model \\
Writing form & A versus B & No detected difference; wide uncertainty \\
Carrier form & Resident versus BM25 in two small batches & No detected
behavioral difference; not equivalent \\
Runtime & Measured first-turn prompts and provider records & Lower first-turn
knowledge-component at $k=5$ in one L1 point measurement; session prompts
$\measured{1.18}\times$; last-turn ranges overlap; dollars $\measured{1.302}\times$
with undecomposed cache pricing \\
Portability & File-copy migration & No code integration or restart required;
behavioral preservation untested \\
Recursive gain & Implemented pathway only & Compounding is unmeasured \\
\bottomrule
\end{tabularx}
\end{table}

\paragraph{Artifact map.}
The behavioral studies live under
\texttt{data/03-pilot/\{semantic-semret-v2,memret,migration\}}. The
availability batch lives in \texttt{data/03-pilot/results/expand-L5-verdict.md}
with records \texttt{data/03-pilot/records/expand-L5-*.jsonl} and the historical
floor \texttt{data/03-pilot/records/calib-step2-floor.jsonl}. Resident-index
instrumentation is under \texttt{data/04-instrumentation}; design audits and the
claim--evidence ledger are under \texttt{design/00-design}; integrated
recomputation is under \texttt{scripts/}. \texttt{MANIFEST.txt} gives the size
and SHA256 of every distributed file. Invalid and superseded attempts remain
in the records and are excluded only according to the frozen protocol.

\paragraph{Disclosure boundary.}
The audited loader treats memory as an ordinary Skill: its
\texttt{description} is resident and its body and references are opened on
demand. The controlled-write gate refuses descriptions above
\measured{1{,}024} characters; an ordinary file edit can bypass that gate and
does not update the running process immediately. The bound is therefore an
enforced property of the governed path, not a universal filesystem invariant.

\section{Behavioral protocols and uncertainty}
\label{sec:pilot}

\subsection{Knowledge-availability floor}

The activating task asks whether an agent reproduces a previously observed L5
failure (missing binding or missing runner confirmation, conjunctive checker).
Lower is better. Form A carries the rule in $L_0$ (with the full
lesson also in $L_2$); Form B carries a narrative in $L_0$ with the full
lesson only in $L_2$; $L_1$ is an identical invite stub in both; C
withholds everything. The valid outcomes are A \measured{0/8}, B \measured{1/8}, and C
\measured{6/6}, all on \texttt{opencode-go/deepseek-flash}. C is a same-day
historical floor, not concurrent with the A/B expansion; the \measured{8/8/6}
allocation reuses six calibration units. The Fisher tables are
$\bigl(\begin{smallmatrix} 0 & 8 \\ 6 & 0 \end{smallmatrix}\bigr)$ (A vs C)
and $\bigl(\begin{smallmatrix} 1 & 7 \\ 6 & 0 \end{smallmatrix}\bigr)$ (B vs C),
rows $=$ arms, columns $=$ (fail, avoid); values are \measured{0.00033} and
\measured{0.00466}, unadjusted for multiplicity across the two tests sharing
the floor arm. A versus B gives a
\measured{-12.5}-percentage-point difference, CI
$[\measured{-47},\measured{+30}]$ pp, and \measured{p=1.0}. The experiment
therefore reports a post-selection availability association on this task and
model, not superior phrasing.

\subsection{Resident versus retrieved carrier}

The first carrier batch crosses five tasks with two paired seeds. Each arm has
\measured{10} valid units and all \measured{20} deliver artifacts. Lesson
application is \measured{8/10} in each arm, a difference of \measured{0} pp
with CI $[\measured{-45.3},\measured{+45.3}]$ pp and Fisher
\measured{p=1.0}. A screened batch admits only tasks failed twice without the
target lesson. Four of five candidates enter; T2 is voided. Among the admitted
units, the descriptive contrast is resident \measured{6/8} versus BM25
\measured{5/8}, difference
\measured{+12.5} pp, CI $[\measured{-35.5},\measured{+57.5}]$ pp, and
McNemar \measured{p=1.0}. The frozen preregistration's literal \S2.1 criterion
would read supported; because T2 was voided, this paper adopts the recomputer's
more conservative insufficient/void. The first batch, paired by lesson and seed,
has paired table $\bigl(\begin{smallmatrix} 8 & 0 \\ 0 & 2 \end{smallmatrix}\bigr)$
(rows $=$ I applied?, columns $=$ R applied?): \measured{8} agree-pass,
\measured{2} agree-fail, \measured{0} discordant (McNemar \measured{p=1.0});
the reported Fisher value and interval are unpaired descriptives. Main-batch intervals subtract Wilson
endpoints (a conservative approximation, not Newcombe/MOVER); the screened
batch quotes its frozen recomputer's interval descriptively only because the
batch is voided (a Wald-type hybrid, valid under no standard method: neither
Wald nor Newcombe score). Both intervals are too wide to establish
equivalence. The first batch runs \texttt{openrouter/z-ai/glm-5.3-flash} and
the screened batch \texttt{openrouter/deepseek/deepseek-v4.1-flash}.

\paragraph{Non-comparable diagnostic.}
\texttt{target\_found} means an observed reference-file read in the resident
arm but inclusion in injected top-$k$ text in the BM25 arm. Its
\measured{3/10} and \measured{8/10} values are reported separately and never
contrasted; neither is a common measure of body delivery before action.
Session transcripts are not archived, so the resident path-substring aperture
cannot be re-scanned; \measured{3/10} is a lower bound.

\paragraph{Retrospective delivery stratification.}
This retrospective proxy tabulation stratifies frozen units on the delivery
proxy (observed reference read in the resident arm; top-$k$ inclusion in the
retrieval arm): \measured{3/3} versus \measured{5/7} resident (Fisher
\measured{p=1.0}) and \measured{6/8} versus \measured{2/2} retrieved
(\measured{p=1.0}); the screened batch gives \measured{2/2} versus
\measured{4/6} resident, with all \measured{8} retrieved units carrying the
target, so prediction cannot be assessed there. No detected association in any
assessable batch (tiny-$n$, task-confounded, all $p{=}1.0$). The resident and retrieval proxies
answer different questions and are reported separately, never contrasted. Both
retrieved lesson-2 units fail despite delivery ($D \not\Rightarrow U$) while
all four lesson-3 units apply without it ($O$ without $D$); both tasks are
retained per
protocol, and their heterogeneity motivates body-necessary task design for the
prospective $h$ study.

\subsection{E1: description-only ablation (preregistered, n=8/arm)}
\label{sec:e1}

\paragraph{Design.}
$D_0$ (description-only, L1 stub) versus $D_1$ (full resident) on the L5
delegation task, with \measured{8} valid units per arm on paired seeds
(\measured{11}/\measured{22}, \measured{4} replicates each, paired by the
seed--replicate key); the L0 text is byte-identical across arms and the only
permitted difference is the L1 body. Both arms run
\texttt{openrouter/deepseek/deepseek-v4.1-flash}; E1 values are juxtaposed with RQ1 qualitatively only, never
tabulated together for statistical comparison.

\paragraph{Main outcome (frozen instruments, descriptive only).}
Fail rates are \measured{0/8} in $D_0$ and \measured{1/8} in $D_1$, a difference
$\Delta = \measured{-12.5}$~pp ($D_0$ minus $D_1$, so positive would mean the body
helps); Fisher exact two-sided \measured{p=1.0} on
$\bigl(\begin{smallmatrix} 0 & 8 \\ 1 & 7 \end{smallmatrix}\bigr)$ (rows
$=$ arms, columns $=$ fail/avoid); Wilson \measured{95\%} intervals
$[\measured{0.00\%},\measured{32.44\%}]$ ($D_0$) and
$[\measured{2.24\%},\measured{47.09\%}]$ ($D_1$), subtracted as
$[\measured{-47.09},\measured{+30.20}]$~pp (endpoint subtraction, a conservative
approximation, not Newcombe/MOVER); paired McNemar exact two-sided
\measured{p=1.0} (\measured{7} concordant-avoid pairs, \measured{0}
concordant-fail, one discordant pair at seed \measured{22} replicate \measured{1});
binding holds in \measured{100\%} of both arms and runner confirmation in
\measured{100\%} ($D_0$) versus \measured{87.5\%} ($D_1$, \measured{7/8}; the
single fail keeps its binding and misses only the runner-confirmation form). At
\measured{n=8} per arm the interval spans on the order of $\pm\measured{45}$~pp,
so every $p$ value and interval in this subsection is descriptive, never
confirmatory. No RQ1 value shares a table with E1 values here.

\paragraph{Prospective $h$.}
The delivery flag was logged prospectively in every unit under one schema
decoupled from the checker: \measured{0/16} body/reference opens before action
across both arms, so the $D{=}1$ stratum is empty in each arm. This is a
prospective zero reading, not a retrospective lower bound, and it is not
comparable with the historical retrospective proxy.

\paragraph{Zone verdict (frozen zone table, provisional).}
The frozen table provides: Z-suffice (description suffices, provisional) when
$\Delta \le \measured{+12.5}$~pp and, within $D_0$, $D{=}1$ versus $D{=}0$ shows
no association (Fisher $p{=}1.0$ direction) with equal $D{=}0$-substrata fail rates
across arms; Z-necessary (body-necessary signal) when $\Delta \ge
\measured{+25}$~pp with gains concentrated in the $D{=}1$ stratum; Z-mixed when
the outcome layer meets a threshold but the stratification fails the
corresponding attached condition; Z-indeterminate otherwise; both named zones are
provisional at this $n$. Here $\Delta = \measured{-12.5}$~pp meets the Z-suffice
outcome layer but not the Z-necessary one, and the attached condition fails: the $D{=}1$
stratum is empty in both arms (the within-$D_0$ comparison degenerates to
\measured{0/0} versus \measured{0/8}, which cannot count as a satisfied
no-association check), and the $D{=}0$ substrata are unequal (\measured{0/8}
versus \measured{1/8}). The verdict is therefore Z-mixed (provisional):
mechanism unclear, outcome difference only, no carrier conclusion.

\paragraph{Plain description (no zone or carrier language).}
\measured{15} of \measured{16} valid units avoid the L5 failure and \measured{1}
fails; no unit opened the skill body or any reference before acting; within the
$D{=}0$ substrata $D_0$ avoids \measured{8/8} and $D_1$ avoids \measured{7/8}, the
single fail being a final-answer citation-form miss while its binding stands.
Literally: the description stayed resident throughout yet the body was never
opened, and the batch still met both conjunctive requirements in all but one
unit.

\paragraph{Sensitivity.}
Dropping the two read-visibility-flagged $D_0$ units at seed \measured{22}
(replicates \measured{2} and \measured{3}) gives \measured{0/6} versus
\measured{1/8} with the same \measured{-12.5}~pp difference and an unchanged
Z-mixed (provisional) verdict.

\paragraph{Disclosures.}
The preregistration froze at \texttt{2026-09-12T14:54:00Z} with a three-item
amendment (sandbox source switched to the archived deployment tree; dual-hash
reconciliation of the memory text matched so the archived copy was used directly
with no deviation; execution-order fix of the setup snippet with the scheme
unchanged); the checker ran at v2 throughout (v1 misfired both ways under this
harness, so v2 recognizes only genuine dispatch lines while the prose rule is
unchanged and v1 stays archived); seeds are newly set with the paired structure
kept; $D_0$ unmounts L1 by file-stub replacement because no first-class unmount
switch exists; the prospective $h$ schema is E1-internal and not comparable with
history; the single-turn sandboxed call chain is newly built with a turn shape not
fully isomorphic to the original; the reconstructed task wording, the strict
before-dispatch binding and read-back checker details, and the stand-in recall
text are disclosed limitations; in-unit re-bindings after credential or allowlist
refusals earned B/R through the unit's own dispatch and read-back; a few units
performed read-only recon of sibling records, traces, or the checker source inside
the run tree with zero writes and stay valid per precedent subject to validity
review; system-temp scratch outside the run and frozen trees, dismissed orphan
first legs, one retained in-file void row with analysis taking the latest valid row
per unit, and the same-key retry rule with its stop criterion never triggered are
all on record; sixteen per-unit records ship with the artifact.

\begin{figure}[t]\centering
\includegraphics[width=0.93\textwidth]{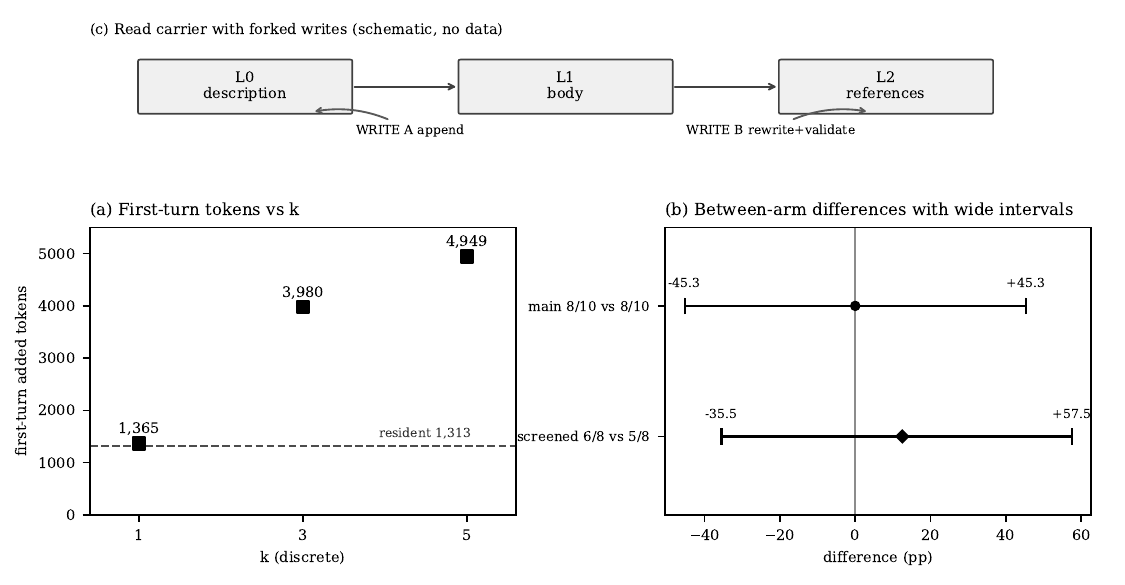}
\caption{Measured operating point, appendix exhibit (same slices and denominators as Tables~\ref{tab:runtime-cost} and~\ref{tab:behavior}). (a) First-turn added tokens at $k=1,3,5$: BM25 \measured{1{,}365}, \measured{3{,}980}, \measured{4{,}949} (discrete points) against resident \measured{1{,}313} (same-unit deltas; absolute baseline in Table~\ref{tab:runtime-cost}). (b) Between-arm application contrasts with wide intervals: \measured{8/10} vs \measured{8/10}; $\Delta=\measured{0}$\,pp, CI $[\measured{-45.3}, \measured{+45.3}]$\,pp; screened \measured{6/8} vs \measured{5/8}, CI $[\measured{-35.5}, \measured{+57.5}]$\,pp. Top strip recalls the $L_0 \to L_1 \to L_2$ read carrier with append versus rewrite--revalidate writes; delivery $h$ is unmeasured in RQ1--RQ2 and instrumented in E1 (\S\ref{sec:e1}).}
\label{fig:e1-data}
\end{figure}

\section{Runtime, migration, and maintenance}
\label{sec:compare}

\subsection{Runtime accounting}

The no-knowledge first-turn prompt is a single L1 point measurement of
\measured{1{,}462} tokens. Subtracting that baseline gives
\measured{1{,}313} tokens for the resident arm and
\measured{1{,}365}, \measured{3{,}980}, and \measured{4{,}949} for BM25 at
$k=1,3,5$. Token-curve sessions ran under a \measured{120}s cap and take
first-turn prompts only; they are marked not-run for behavior, and $k=0$ is a
token baseline rather than a behavior arm. The behavioral comparison uses $k=5$.
Across its valid units, provider-reported cost is \measured{\$0.09619} resident
versus \measured{\$0.12521} retrieved; prompt volume and provider pricing are not
interchangeable, so the paper reports both rather than inferring a price
decomposition. Session prompt totals are \measured{1{,}218{,}713} versus
\measured{1{,}436{,}897} ($\measured{1.18}\times$); uncached input differs by
$\measured{1.84}\times$ while cache-read tokens differ by only
$\measured{1.05}\times$, so cache pricing moves the dollar ratio independently
of prompt size. The calibrated char-to-token interval for the resident index
does not contain the measured first-turn delta, so char counts are never used
as token estimators here. Last-turn prompt ranges overlap
(\measured{5{,}181}--\measured{29{,}955}
versus \measured{9{,}398}--\measured{30{,}517}). A second batch
(semantic-semret-v2, voided; descriptive only) gives mean first-turn
\measured{+1{,}522.8} resident versus \measured{+11{,}999.2} BM25 at $k=5$
($\sim\measured{7.9}\times$): the ratio is lesson- and corpus-dependent.
Cite $\measured{3.8}\times$ as the L1 first-turn point measurement;
$\sim\measured{7.9}\times$ is the voided batch's mean, reported to bound
regime-dependence.

\begin{table}[h]
\centering\small
\caption{What differs between the carrier arms. Task, model, and paired seeds
are held fixed.}
\begin{tabularx}{\textwidth}{@{}p{2.5cm}p{2.7cm}X@{}}
\toprule
Dimension & Resident carrier & BM25 carrier \\
\midrule
Routing & Resident descriptions; body read on demand & Query and inject top-$k$
body text \\
Knowledge context & Fixed compact index each turn & Retrieved text injected
each turn \\
External retrieval call & None in the agent loop & BM25 selection before the
model call \\
Irrelevant carriers & Five unrelated skills also resident & Zero skills \\
Behavioral endpoint & \measured{8/10} applied & \measured{8/10} applied \\
\bottomrule
\end{tabularx}
\end{table}

The arms differ by more than delivery: the resident arm mounts six real skills
while the retrieval arm mounts zero, so the comparison is a deployed system
against a retrieval arm. Injection is unconditional every turn, matching the
batch's every-turn cost shape; a conditional-retrieval policy would shift the
profile, and the $k=1$ near-tie bounds the headline.

This is a comparison with sparse BM25 retrieval, not with dense, graph, or
learned retrieval. At $k=1$ the measured context costs are nearly tied; at
$k=5$ the resident arm is smaller. Corpus size, prompt caching, pricing, and
semantic-retrieval coverage requirements can move the cost comparison.

\subsection{Portability and maintenance}

Migration starts by copying \measured{21} files and \measured{177{,}321} bytes
(manifest snapshot, taken as authoritative); a second snapshot records
\measured{185{,}582} bytes, and the recording file internally contradicts
itself about which snapshot it describes, so no drift-over-time claim is made.
The migration changes \measured{0} code lines and restarts \measured{0}
services. Carriers are
present in all \measured{8/8} inspected units. The behavioral endpoint is
\measured{0/8} versus \measured{0/8} with \measured{p=1.0}; this double floor
cannot establish preservation.

Across the observed maintenance window, the content store grows
\measured{+64\%}; none of \measured{11} content changes touches harness code.
The machinery itself changes \measured{15} times (\measured{13}
instrumentation plus \measured{2} write-entry, per the pinned deployment
history's changed paths). The supported conclusion is
that content addition was decoupled from harness edits in this window, not that
the machinery is free or asymptotically constant. Migration units run
\texttt{openrouter/z-ai/glm-5.3-flash}.

\section{Write-governance audit}
\label{sec:audit}

The installed system exposes four physical write entries
(Table~\ref{tab:appendix-writes}). The official carrier operation can rewrite a
\texttt{SKILL.md} or append a reference attachment after validation, but three
other entries remain. Only \texttt{apply\_change} records an expectation in the
settlement ledger; coverage is therefore \measured{1/4}.

\begin{table}[h]
\centering\small
\caption{Physical write surface on the audited deployment. Source locations
and excerpts are in \texttt{design/00-design/live-citations.txt}; the 15-edit
count traces to deployment history at commit \texttt{57ecad8dd3}.}
\label{tab:appendix-writes}
\begin{tabularx}{\textwidth}{@{}p{2.65cm}p{2.55cm}X@{}}
\toprule
Entry & State transition & Governance \\
\midrule
\texttt{apply\_change} & Rewrite skill or append attachment & Gate, adoption,
rollback, settlement \\
\texttt{facts\_append} & Append raw evidence & Append enforced by construction;
no settlement \\
\texttt{lesson\_append} & Append lesson detail & Local refusal rule; no
settlement \\
Bare write/edit & Arbitrary file mutation & Bypasses carrier gate and ledger \\
\bottomrule
\end{tabularx}
\end{table}

The official carrier operation now routes reference attachments to a
path-confined, append-only target. This closes official reachability but does
not remove the other entries or extend ledger coverage. The architectural
distinction is semantic: evidence history is append-only, whereas current
executable state is rewritable and must be revalidated.

\section{Verification and reproduction}
\label{sec:reflexive}
\label{sec:verifyself}
\label{sec:repro}

\paragraph{Recompute.}
From the artifact root (Python $\ge 3.12$):
\begin{quote}\footnotesize\raggedright
\texttt{python3 scripts/section7\_recompute.py --json --no-checker}\\
\texttt{python3 scripts/recompute\_rq1\_sessions.py --json}\\
\texttt{python3 data/03-pilot/memret/analyze\_memret.py}\\
\texttt{python3 data/03-pilot/migration/analyze\_migration.py}\\
\texttt{python3 data/03-pilot/semantic-semret-v2/recompute\_v2.py --json}
\end{quote}
All but the last exit \measured{0} (migration also writes its
\texttt{analysis.json}). The final command exits \measured{4} by design because
the preregistered screened batch is marked insufficient/void; it still
reproduces its descriptive statistics.
All \measured{20/20} memret answers match the unit ledger on byte length, and
the five frozen checkers reproduce all recorded artifact decisions.

\paragraph{Next measurements.}
Five instruments would each unlock a decision this study cannot make:
(i) $h$, separating ``the carrier failed'' from ``the carrier never
delivered''---now instrumented prospectively in E1 with \measured{0/16}
body/reference opens, so the missing measurement is a zero reading, not a gap;
(ii) an equivalence design with a predeclared margin, licensing
``no worse than retrieval''; (iii) a dense-retrieval arm, locating the
crossover beyond sparse BM25; (iv) a non-floor migration endpoint, testing
behavioral preservation; (v) repeated reflection cycles, testing compounding
gain. These are missing measurements, not zero-valued results;
\S\ref{msec:discussion} orders them by information value. A frozen
preregistration for the cheapest decisive test---a description-only ablation
on the L5 task, $n=8$/arm on paired seeds---ships in
\texttt{design/00-design/prereg-e1-description-only.md}.

\paragraph{Provenance and ethics.}
Figure~\ref{fig:carrier} carries the implemented loop over an AI-generated backdrop (labels overlaid). Figure~\ref{fig:e1-data} was generated by \texttt{figs/data-panels.py} from \texttt{figs/data-values.json}; all plotted values repeat Tables~\ref{tab:runtime-cost} and~\ref{tab:behavior} verbatim. Drafting used LLM
assistance with human verification of every reported number. The study uses no
human subjects and no personal data.

\end{document}